\documentclass[3p,preview]{elsarticle}
\usepackage{multirow}
\usepackage[table,xcdraw]{xcolor}
\usepackage{soul}
\usepackage{float}
\usepackage{amssymb,amsmath}
\usepackage{graphicx}
\usepackage{color}
\usepackage[textwidth=2.5cm]{todonotes}    

\usepackage{lineno,hyperref}
\modulolinenumbers[5]

\journal{Computational Statistics \& Data Analysis}

\begin{document}

\begin{frontmatter}

%%%%%%%%%%%%%%%%%%%%%%%%%%%%%%%%%%%%%%
%%%%%   TITLE

\title{Band Depth based initialization of $k$-Means for functional data clustering}

%% Authors:
\author[stat]{Javier Albert--Smet}

\author[math]{Aurora Torrente\corref{mycorrespondingauthor}}
\cortext[mycorrespondingauthor]{Corresponding author}
\ead{etorrent@est-econ.uc3m.es}

\author[stat]{Juan Romo}

\address[stat]{Universidad Carlos III de Madrid, Departamento de Estad\'istica, C/ Madrid 126, 28903  Getafe, Madrid, Spain }
\address[math]{Universidad Carlos III de Madrid, Instituto Gregorio Mill\'an, Av. Universidad 30, 28911 Legan\'es, Madrid, Spain}

\nonumnote{Declarations of interest: none.}

%%%%%%%%%%%%%%%%%%%%%%%%%%%%%%%%%
%%%%%     ABSTRACT

\begin{abstract}

\hyphenation{FABRIk}

The $k$-Means algorithm is one of the most popular choices for clustering data but is well-known to be sensitive to the initialization process. There is a substantial number of methods that aim at finding optimal initial seeds for $k$-Means, though none of them are universally valid. This paper presents an extension to longitudinal data of one of such methods, the BRIk algorithm, that relies on clustering a set of centroids derived from bootstrap replicates of the data and on the use of the versatile Modified Band Depth. In our approach we improve the BRIk method by adding a step where we fit appropriate B-splines to our observations and a resampling process that allows computational feasibility and handling issues such as noise or missing data. Our results with simulated and real data sets indicate that our \textit{F}unctional Data \textit{A}pproach to the BRIK method (FABRIk) is more effective than previous proposals at providing seeds to initialize $k$-Means in terms of clustering recovery. 

\end{abstract}


\end{frontmatter}
\newpage

%\linenumbers

%%%%%%%%%%%%%%%%%%%%%%%%%%%%%%%%%
%%%%%     INTRODUCTION

\section{Introduction}
\label{intro}

Amongst all non-hierarchical clustering algorithms, \textit{k}-Means is  the most widely used in every research field, from signal processing to molecular genetics. It is an iterative method that works by allocating each data point to the cluster with nearest gravity center until assignments no longer change or a maximum number of iterations is reached. Despite having a fast convergence to a minimum of the distortion --i.e., the sum of squared Euclidean distances between each data point and its nearest cluster center-- \citep{Selim:84}, the method has well known disadvantages, including its dependence on the initialization process. If inappropriate initial points are chosen, the method can exhibit drawbacks such as getting stuck on a bad local minimum, converging more slowly or producing empty clusters \citep{Celebi:11}. The existence of multiple local optima, which has proven to depend on the dataset size and on the overlap of clusters, greatly influences the performance of $k$-Means \citep{Steinley:06}. Also, the method is known to be NP-hard \citep{Garey:79}; this has motivated numerous efforts to find techniques that provide sub-optimal solutions. Therefore, there  are several methods for initializing \textit{k}-Means with suitable seeds, though none of them universally accepted;  see \cite{He:04, Steinley:07, Celebi:13}, for example.

Recently, the BRIk method (Bootstrap Random Initialization for \textit{k}-Means) has been proposed in \cite{BRIK} as a relevant alternative. This technique has two separate stages. In the first one, the input dataset is bootstrapped $B$ times. \textit{k}-Means is then run over these bootstrap replicates, with randomly chosen initial seeds, to obtain a set of cluster centers that form more compact groups than those in the original dataset. In the second stage these cluster centers are partitioned into groups and the deepest point of each grouping is calculated. These prototypes are used as the initialization points for the \textit{k}-Means clustering algorithm.

The BRIk method is flexible as it allows the user to make different choices regarding the number $B$ of bootstrap replicates, the technique to cluster the bootstrap centers and the depth notion used to find the most representative seed for each cluster. 
There is a variety of data depth definitions, all of which allow generalizing the unidimensional median and rank to the multivariate or the functional data contexts. The main difference between them is that the latter models longitudinal data as continuous functions and thus incorporates much more information into the analysis. Over the last two decades there has been a substantial  growth in the functional data analysis (FDA) literature, including techniques for cluster analysis \citep{Ferreira:09, Jacques:13}, classification \citep{Sara:03, Leng:06}, analysis of variance \citep{Zhang:13} and principal components analysis \citep{Hall:18}, among others. In particular, the  --computationally feasible-- Modified Band Depth (MBD) \citep{Sara:09} has been successfully applied to FDA for shape outlier detection \citep{arribas:14}, functional boxplot construction \citep{sun:11} or time warping \citep{arribas:11}; also, for BRIk, it is recommended to make use of the multivariate version of the MBD.

In this work we consider the FDA context and propose an extension of BRIk, that we have called the Functional data Approach to BRIk (FABRIk) method. The underlying idea is simple. We fit a continuous function to longitudinal data and sample a number $D$ of time points, which will be clustered using $D$-dimensional seeds, thus providing the final output labels. This offers computational feasibility and several advantages over standard multivariate techniques, including the possibility of smoothing data to reduce noise or including observations with missing features (time points) into the analysis.

The paper is organized as follows. Section~\ref{methods} describes in detail our algorithm and the methods it will be compared to. Section~\ref{experimental} specifies the data, both simulated and real, and the quality measures used to assess FABRIk. Section~\ref{section.results} presents the overall results and in Section~\ref{section.discussion} we summarize our findings.

%%%%%%%%%%%%%%%%%%%%%%%%%%%%%%%%%%%%%%%%%%%%%%%%%%%%%%%%%%%%%%%%
%%%%%                METHODS  OVERVIEW
%%%%%%%%%%%%%%%%%%%%%%%%%%%%%%%%%%%%%%%%%%%%%%%%%%%%%%%%%%%%%%%%

\section{Methods}\label{methods}

Consider a multivariate dataset $X=\{x_1,\dots,x_N\}$ of $N$ observations in $d$ dimension, $x_i=(x_{i1}, \dots , x_{id})$, $i=1, 2, ... , N,$ that have to be partitioned into $K$ clusters $G_1,\dots,G_K$ by means of the \textit{k}-Means algorithm \citep{Forgy:65}. Once $K$ initial seeds (centers) $c_1,\dots,c_K$ have been obtained, \textit{k}-Means works in the following way. Each data point $x_i$ is assigned to exactly one of the clusters, $G_{i_0}$, where $\displaystyle i_0 = \arg \min_{1 \le j \le K} d(x_i,c_j)$. Next, for the $j$-th cluster, the centers are updated by calculating the component-wise mean of elements in $G_j$. The assignment of elements to clusters and the computation of centers are repeated until there are no changes in the assignment or a maximum number of iterations is reached.

Since the $k$-Means output strongly depends on the initial seeds, there have been numerous efforts in the literature to obtain suitable initial centers for \textit{k}-Means. In this work we focus on extending one of these methods, the BRIk algorithm, summarized as follows.

\begin{itemize}
 \item[S1.] FOR ($b$ in $1:B$) DO
  \begin{itemize}
   \item Obtain the $b$-th bootstrap sample $X_b$ of the set $X$. 
   \item Run $k$-Means, randomly initialized with $K$ seeds, on $X_b$ and store the final $K$ centroids. 
  \end{itemize}
 \item[S2.] Group the dataset of $K \times B$ centroids into $K$ clusters, using some non-hierarchical algorithm.
 \item[S3.] Find the deepest point of each cluster from step S2, using some data depth notion; these are used as initial seeds of $k$-Means.
\end{itemize}

Specifically, it is recommended to use the MBD depth notion. Thus, the BRIk algorithm's third step relies on finding the MBD-deepest point of each cluster (center grouping) from S2. The method is designed to use any clustering technique in this step. Here we used the Partitioning Around the Medoids (PAM) method \citep{Kaufman:90}. Note that the Ward algorithm \citep{Ward:63} also reported a good performance in the experiments \cite{BRIK}.

The method that we present here is an extension of BRIk. In the case of data that come from function observations we can add another stage, a functional approximation, to enhance the behavior of the method. In particular, we take the B-splines that best fit the original data in the least squares sense. This process sets a basis of (continuous) piecewise polynomial functions of a given degree and constructs the linear combination of these that best fits the data, to provide an approximation to the original function that is continuous and differentiable to a certain order.

 The use of MBD with continuous functions is straight forward. Given a set of $N$ real functions $x_1,\dots,x_N$ that are continuous on the compact interval $I$, the MBD of a function $x$ within such a set is given by 
\begin{equation*}
    MBD(x) = \binom{n}{j}^{-1}  \sum_{1 \le i_1 < i_2 \le N}  \lambda_r(A(x;x_{i_1}, x_{i_2})),
\end{equation*}
where $A(x)\equiv A(x;x_{i_1}, x_{i_2})=\{t\in I \colon \min_{r=i_1,i_2} x_r(t) \le x(t) \le \max_{r=i_1,i_2} x_r(t)\}$ and $\lambda_r(A(x)) = \lambda(A)/\lambda(I)$, and $\lambda$ is the Lebesgue measure on $I$. Thus, $\lambda_r(A(x))$ represents the proportion of time that $x$ is in the band defined by functions $x_{i_1}$ and $x_{i_2}$.

With this approach, we have all the advantages of continuous functions. From a computational perspective we can evaluate the functions obtained on new time points (on a grid as dense as desired) to get a new dataset in $D$ dimension. Then, this re-sampled data are input to the BRIk algorithm to find the initial seeds of \textit{k}-Means.

To check the performance of our method we have selected, as benchmark, the classical Forgy approach \citep{Forgy:65}, where the initial seeds are selected at random; we refer to this as the KM initialization. 

Next, we have considered a widely-used algorithm, \textit{k}-Means++ (KMPP) \citep{Arthur:07}, which aims at improving the random selection of the initial seeds in the following way. A random data point $c_1$ is firstly picked from the data set. This 
conditions the selection of the remaining initial seeds $c_j$, $j=2,\dots,K$, which are sequentially chosen among the remaining observations $x$ according to a probability proportional to $d^2(x, c_{j_x})$, the squared distance between the point $x$ and its closest seed $c_{j_x}$, where $j_x = \displaystyle \arg \min_{1\le i < j - 1} d(x, c_i)$. Following this procedure, the initial centers are typically separated from each other and yield more accurate groupings.

Additionally, in order to assess the potential improvement of our method, the functional approximation stage of the  FABRIk method is added before the KM and the KMPP method are run. This way we can make a complete and fair comparison of how the FDA approach affects the BRIk method against how it improves KM or KMPP.

Hence, on the one hand, we will compare six different \textit{k}-Means initialization techniques: KM and its FDA version, designated as FKM, BRIk,  FABRIk, KMPP and KMPP with the functional approximation (denoted by FKMPP). On the other hand, we want to compare our strategy of replacing the $d$-dimensional data by those estimated in $D$ dimension against the popular proposal of clustering the B-splines coefficients \citep{Abraham:03}. This is because two functions in the same cluster are expected to have similar vectors of coefficients, and also because in most situations these vectors are considerably smaller in size than the $D$-dimensional ones. Therefore, we ran $k$-Means on the set of computed coefficients, initialized with KM, BRIk and KMPP. We will refer to these approaches as C-KM, C-BRIk and C-KMPP.

In order to explore the advantages of our method, we have also carried out experiments where the data points had missing observations, what translates into ''sparse data'' in the functional data context. For each simulated dataset we randomly removed a proportion $p\in\{0.1, 0.25, 0.5\}$ of the coordinates of each $d$-dimensional vector. For FKM, FABRIk and FKMPP, we estimated each missing value in a given vector by means of the corresponding B-spline. For KM, BRIk and KMPP, we imputed the missing data by using linear interpolation; note that, for simplicity, the removal of coordinates was hence not applied to the first and last values. We then performed the analysis of the resulting data with each of the nine methods mentioned.

%%%%%%%%%%%%%%%%%%%%%%%%%%%%%%%%%
%%%%%%%
%%%%%%%				RESULTS
%%%%%%%
%%%%%%%%%%%%%%%%%%%%%%%%%%%%%%%%%

%%%%%%%%%%%    SETUP

\section{Experimental Setup}\label{experimental}
 
Our experiments were carried out in the R statistical language \citep{Rproject}, using the implementation of $k$-Means included in the \verb|stats| package, that of B-splines in the \verb|splines| package  and the $MBD$ coding provided in the \verb|depthTools| package \citep{BMC:13}. For each dataset, the number of clusters in the $k$-Means algorithm was set to be equal to the number of groups. For  FABRIk and BRIk, we used bootstrap sizes of $B=25$. Using a larger bootstrap size decreases the speed of the  FABRIk and BRIk algorithms, while slightly improving the distortion. Cubic B-splines with no intercept and a varying number of equally spaced knots, depending on the model to be analyzed, were chosen to approximate our data; then new evenly spaced observations were obtained by using an oversampling process with different oversampling factors. An oversampling factor of $m$ means that the number $D$ of time points observed in the approximated function is $m$ times the number of original input samples: $D=m\times d$. The knots are defined through the degrees of freedom (DF) parameter. A DF value of $n$ with cubic B-splines implementation means that $n-3$ internal knots are placed uniformly in the horizontal axis. The resemblance of the approximated function to the real one in each of the models is determined by the DF parameter of the B-splines.

\subsection{Datasets}

\begin{figure}[!h]
  \begin{center}
    \includegraphics[width=\textwidth, trim=0 1mm 0 0, clip]{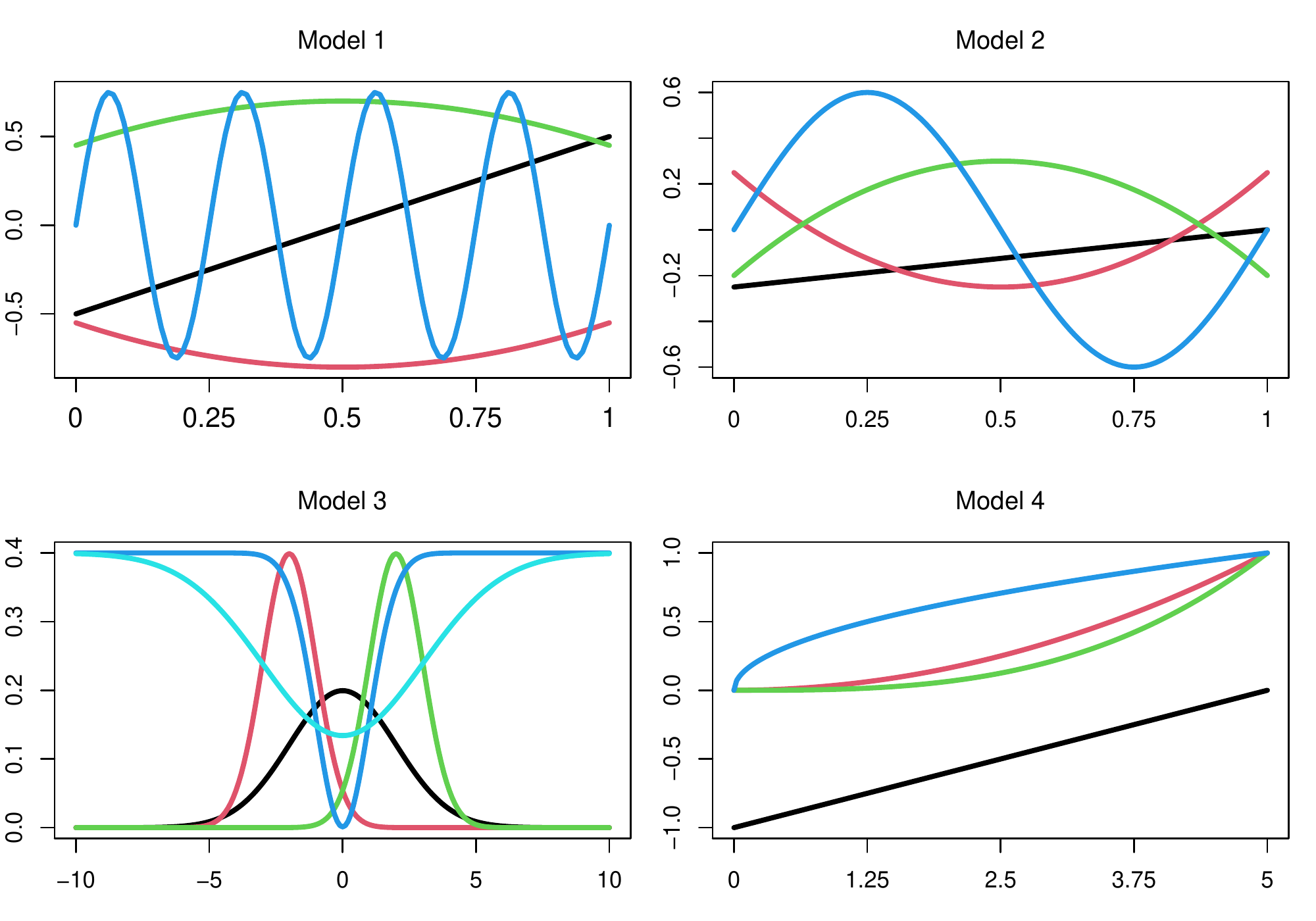}
    \caption{Functions originating each of the clusters for the simulated models by adding Gaussian noise to each sampled component independently.}
    \label{fig1}
  \end{center}
\end{figure}

We conducted experiments involving simulated and real datasets. For the simulated ones we chose the four models described in Table~\ref{modelTab}; the functions giving origin to each of the clusters are shown in Figure~\ref{fig1}. Models 1 and 2 consider polynomial and sinusoidal functions; the former is designed to assess the effect of rapidly changing signals on the clustering quality whereas the latter could be used, for instance, to mimic monthly average temperatures in different climates.  Model 3 consists of (raw and transformed) Gaussian functions and is used to test the impact of sudden peaks on signal clustering. Finally, Model 4, taken from \cite{Leroy:18} attempts to model swimmers' progression curves. The time vector ($x$ coordinate) varies from model to model, while the number of simulated functions per cluster is 25 for all of them. To construct the clusters, additive white Gaussian noise is incorporated to each model to mimic the randomness in the data collection process.

\begin{table}[!htb]
\begin{center}
\begin{tabular}{c|c|c|}
\hline
\multicolumn{1}{|c|}{
\begin{tabular}[c]{@{}c@{}}
  \\[-4mm]
  %%%%%%%%%%%%%%%%%%%%%%%%  Model 1
  Model 1 \\[3mm] 
  (DF $=15$)
 \end{tabular}
} & $x = (0, 0.01, 0.02, ... , 1)$    & 
\begin{tabular}[c]{@{}c@{}}
$y_1 = x-0.5$\\[2mm] 
$y_2 = (x-0.5)^2-0.8$\\[2mm] 
$y_3 = -(x-0.5)^2+0.7$\\[2mm] 
$y_4 = 0.75 \cdot \sin(8 \pi \cdot x)$\\[1mm]
\end{tabular}       
\\ \hline
  %%%%%%%%%%%%%%%%%%%%%%%%  Model 2
\multicolumn{1}{|c|}{ 
\begin{tabular}[c]{@{}c@{}}
  \\[-4mm]
  Model 2 \\[3mm] 
  (DF $=4$)
 \end{tabular}
 } & $x = (0, 0.01, 0.02, ... , 1)$    
 & \begin{tabular}[c]{@{}c@{}}
 $y_1 = x$\\[2mm] 
 $y_2 = 2\cdot(x-0.5)^2-0.25$\\[2mm] 
 $y_3 = -2\cdot(x-0.5)^2+0.3$\\[2mm] 
 $y_4 = 0.6 \cdot \sin(2 \pi \cdot x-0.5)$ \\[1mm]
\end{tabular}  
 \\ \hline
  %%%%%%%%%%%%%%%%%%%%%%%%  Model 3
\multicolumn{1}{|c|}{
\begin{tabular}[c]{@{}c@{}}
  \\[-4mm]
  Model 3 \\[3mm] 
  (DF $=13$)
 \end{tabular}
} & $x = (-10, -9.9, -9.8, ... , 10)$ & 
 \begin{tabular}[c]{@{}c@{}}
  \\[-3mm]
  $y_1 = \dfrac{1}{2\sqrt{2\pi}}\cdot e^{-\frac{(x)^2}{2 \cdot 2^2}}$\\[3mm] 
  $y_2 = \dfrac{1}{\sqrt{2\pi}}\cdot e^{-\frac{(x+2)^2}{2 \cdot 1^2}}$\\[3mm] 
  $y_3 = \dfrac{1}{\sqrt{2\pi}}\cdot e^{-\frac{(x-2)^2}{2 \cdot 1^2}}$\\[3mm] 
  $y_4 = \dfrac{-1}{\sqrt{2\pi}}\cdot e^{-\frac{(x)^2}{2 \cdot 1^2}}+0.4$\\[3mm]
  $y_5 = \dfrac{-2}{3\sqrt{2\pi}}\cdot e^{-\frac{(x)^2}{2 \cdot 3^2}}+0.4$\\[3mm]
 \end{tabular} 
 \\   \hline
   %%%%%%%%%%%%%%%%%%%%%%%%  Model 4
\multicolumn{1}{|c|}{
\begin{tabular}[c]{@{}c@{}}
  \\[-4mm]
  Model 4 \\[3mm] 
  (DF $=4$)
 \end{tabular}
} & $x = (0, 0.05, 0.1, ... , 1)$     & 
 \begin{tabular}[c]{@{}c@{}}
  $y_1 = x-1$\\[2mm] 
  $y_2 = x^2$\\[2mm] 
  $y_3 = x^3$\\[2mm] 
  $y_4 = \sqrt{x}$\\[1mm]
 \end{tabular} 
 \\     \hline
\end{tabular}
\end{center}
\caption{Description of the simulated models. The first column includes the DFs used in each model, according to an elbow-like rule. The second column provides the time vector where the functions are observed; the third column describes the signal defining each cluster. }
\label{modelTab}
\end{table}

The DF parameter requires a careful choice for each model. Higher values of this parameter can account for larger variations of a function, and therefore Models 2 and 3 would require higher DFs. In our experiments, the specific value for each situation was selected in the set \{4,\dots, 50\} according to an elbow-like rule for the plot of the (average) distortion against the DFs; these are provided in the last column of Table~\ref{modelTab}.

For each model we generated 1000 independent datasets that were clustered with the nine methods and considered different levels of noise, $\sigma \in \{0.5, 1, 1.5, 2 \}$. The results reported in this manuscript correspond to $\sigma = 1$. Other values of $\sigma$ produced similar relative outputs; note that increasing the standard deviation to a value greater than $\sigma = 2$ renders a very poor cluster accuracy for every method tested; however, FABRIk and FKMPP present slightly higher accuracy measures than the alternatives as the functional stage is noise-smoothing.

%%%%%%%%%%%%%%%%%%%%%%%%%%%%%%%%%%%%%%%%%%%%%%%%%%%%%%%%%%   REAL    %%%%%%%%%%%%%%%%%%%%%%%%%%%%%%%%%%%%%%%%%%%%%%%%%%%%%%%%%%%%%%%%%%%%%%%%%%% 

To complete the study of our algorithm we used real data to assess whether it is of practical use. 

First, we have considered a dataset containing 200 electrocardiogram (ECG) signals,  by a single electrode, 133 labeled as normal and 67 as abnormal (myocardial infarction), as formatted in \cite{Olszewski:01} (units not provided). Each observation reflects a heartbeat and consists of 96 measurements. The dataset is available at the UCR Time Series Classification Archive \cite{UCRArchive2018}.

%%%% Gyroscope data
Secondly, the Gyroscope dataset was recorded using a Xiaomi Pocophone F1. The mobile phone was laid on a table and moved to follow four patterns: a straight line, a curvy line, an arch and a rotation from side to side on the spot. The yaw angular velocity in $rad/s$ was recorded using the Sensor Record app available in Google Play Store. 

Each recording for each pattern was truncated to 527 registered time points, spaced by 10ms, in order for all data points to have the same length. Thus, their duration is approximately 5 seconds. The dataset consists of 11 recordings for each pattern and is available as Supplementary material.

 Since all the methods tested in this study are random, in the sense that different runs produce in general distinct centroids, we ran each of them 1000 times for each dataset.

\begin{figure}[!h]
  \begin{center}
    \includegraphics[width=\textwidth, trim=0 1mm 0 0, clip]{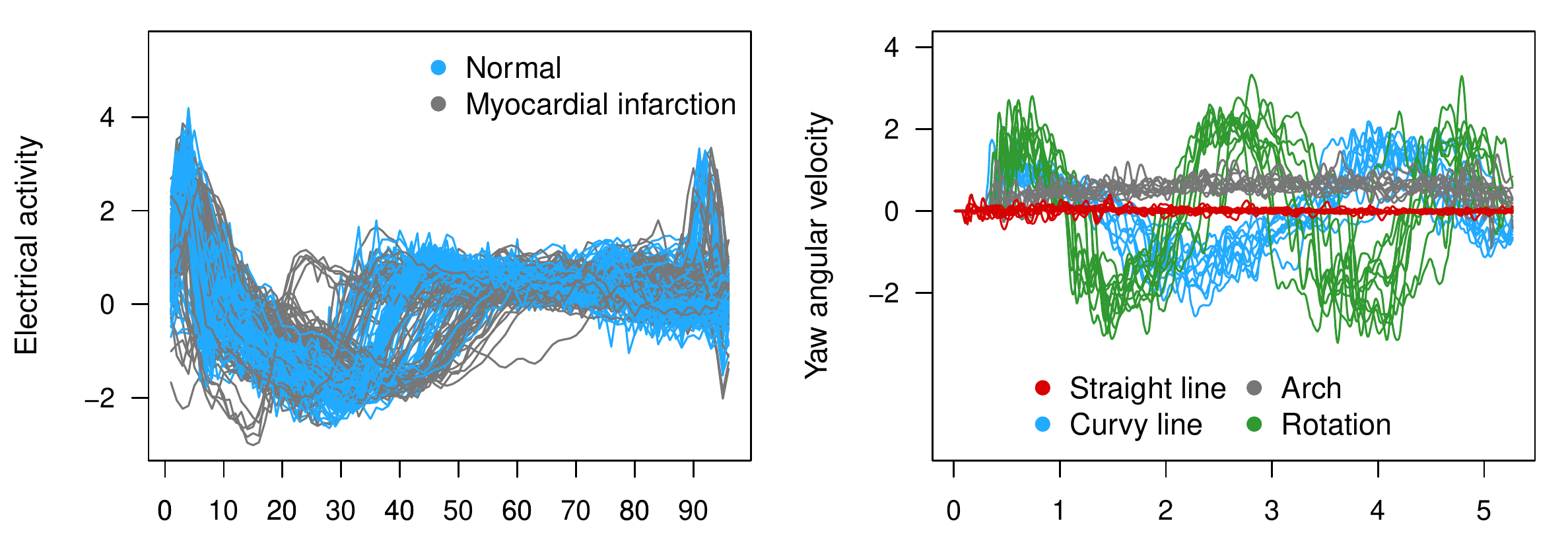}
    \caption{Real data. Left panel: heart electrical activity recorded during a cardiac cycle for patients with a normal heart or with a cardiac condition; units not provided. Right panel: gyroscope yaw velocity readings  (in $rad/s$) for four different patterns, registered at steps of $0.01s$}.
    \label{fig2}
  \end{center}
\end{figure}

%%%%%%%%%%%%%    EVALUATION CRITERIA

\subsection{Performance evaluation}
The overall performance of these methods has been evaluated according to five different measures that fall into four categories: 
\begin{itemize}
    \item Accuracy: 
    We measure how similar the clusters are to the true groups by means of the Adjusted Rand Index (ARI) \citep{Hubert:85} and the clustering correctness, which is computed as the percentage of label agreement (i.e. correctly assigned elements), according to the label permutation that yields the maximum set similarity. 
    \item Dispersion: The obvious choice to determine how compact the clusters $G_1,\dots,G_K$ are is the distortion $\displaystyle\sum_{j=1}^K \sum_{x_i \in G_j} d^2(x_i,c_j)$, where $c_j$ is the gravity center of cluster $G_j$. Its evaluation is done by identifying each cluster from the partitioning labels and     calculating the corresponding centroid, in the original (multivariate) data space. 
    \item Convergence: We assess the convergence speed with the number of iterations required by the \textit{k}-Means algorithm to converge after being initialized.
    \item Computational cost: Finally, we  consider the execution time, in seconds, used by each algorithm from start to finish. Calculations are carried out with an Intel Core i7-6700HQ CPU with 2.60GHz and 8 GB RAM.
\end{itemize}

The performance of the methods we consider is assessed in terms of the median $\tilde{x}$, the mean $\bar{x}$ and the standard deviation $s$ for all five measures.

%%%%%%%%%%%%%%%%%%%%%%%%%%%%%%%%%
%%%%%%%%    RESULTS     %%%%%%%%%
%%%%%%%%%%%%%%%%%%%%%%%%%%%%%%%%%

\section{Results}\label{section.results}

%%%%%%%%%%%%%%%%%%%%%%%%%%%%%%%%%%%%%%%%%%%%%%%%%%%%%%%%%%%%%%
%%%%%%%%%%%             1. SIMULATED DATA        %%%%%%%%%%%%%
%%%%%%%%%%%%%%%%%%%%%%%%%%%%%%%%%%%%%%%%%%%%%%%%%%%%%%%%%%%%%%

\paragraph{Simulated data} We used the four models to evaluate the performance of FABRIk in different situations. For Model 1 the FABRIk method --followed by BRIk-- outperforms the alternatives with respect to all the evaluation measures except for the execution time, as shown in Table~\ref{model1estimates}, where  statistics for the distortion have been rounded to four significant figures. In particular, all the techniques based on clustering the vectors of coefficients are drastically worse than the other ones. 
The same situation is observed for all the scenarios we have considered (different models and levels of noise and presence or absence of missing data) and thus we do not include them in the subsequent tables for compactness.

%%%%%%%%%%%%%%%%%%%%%%%%%%%%%%%%%%
%%%%%%%%%    MODEL 1     %%%%%%%%%
%%%%%%%%%%%%%%%%%%%%%%%%%%%%%%%%%%

\begin{table}[!htb]
\begin{center}
\begin{tabular}{ccccccc}
                        &             & Correctness         & ARI           & Distortion  & Iterations   & Exec. Time (s)  \\ \hline
\multirow{3}{*}{KM}     & $\tilde{x}$ &\textbf{1.0000}      &\textbf{1.0000}& 9744        & 2.000        &\textbf{0.0020}\\
                        & $\bar{x}$   & 0.9236              & 0.9137        & 9799        & 2.518        &\textbf{0.0026}\\
                        & $s$         & 0.1428              & 0.1575        & 265.2       & 0.557        & 0.0027         \\ \hline
\multirow{3}{*}{KMPP}   & $\tilde{x}$ &\textbf{1.0000}      &\textbf{1.0000}& 9755        & 2.000        & 0.0060        \\
                        & $\bar{x}$   & 0.9100              & 0.8986        & 9821        & 2.455        & 0.0066         \\
                        & $s$         & 0.1523              & 0.1681        & 276.2       & 0.542        & 0.0063        \\ \hline
\multirow{3}{*}{BRIk}   & $\tilde{x}$ &\textbf{1.0000}      &\textbf{1.0000}&\textbf{9683}&\textbf{1.000}& 0.1021          \\
                        & $\bar{x}$   & 0.9983              & 0.9961        &\textbf{9687}& 1.301        & 0.1048          \\
                        & $s$         & 0.0120              & 0.0156        & 145.1       & 0.465        & 0.0198         \\ \hline
\multirow{3}{*}{FKM}    & $\tilde{x}$ &\textbf{1.0000}      &\textbf{1.0000}& 9763        & 2.000        & 0.1533          \\
                        & $\bar{x}$   & 0.8996              & 0.8886        & 9852        & 2.247        & 0.1603          \\ 
                        & $s$         & 0.1619              & 0.1787        & 308.4       & 0.550        & 0.0328          \\ \hline
\multirow{3}{*}{FKMPP}  & $\tilde{x}$ &\textbf{1.0000}      &\textbf{1.0000}& 9733        & 2.000        & 0.1956          \\
                        & $\bar{x}$   & 0.9340              & 0.9269        & 9786        & 2.031        & 0.1967          \\ 
                        & $s$         & 0.1355              & 0.1470        & 254.5       & 0.530        & 0.0204          \\ \hline
\multirow{3}{*}{FABRIk} & $\tilde{x}$ &\textbf{1.0000}      &\textbf{1.0000}& 9684        &\textbf{1.000}& 0.2908          \\
                        & $\bar{x}$   &\textbf{0.9992}      &\textbf{0.9977}&\textbf{9687}&\textbf{1.034}& 0.2916          \\
                        & $s$         & 0.0029              & 0.0077        & 144.1       & 0.181        & 0.0226         \\ \hline
\multirow{3}{*}{C-KM}   & $\tilde{x}$ & 0.7700              & 0.6268        & 10290       & 3.000        & 0.1312     \\
                        & $\bar{x}$   & 0.7741              & 0.6299        & 10300       & 3.155        & 0.1411     \\
                        & $s$         & 0.0791              & 0.0924        & 252.4       & 0.676        & 0.0370     \\ \hline
\multirow{3}{*}{C-KMPP} & $\tilde{x}$ & 0.7700              & 0.6272        & 10290       & 3.000        & 0.1368     \\
                        & $\bar{x}$   & 0.7750              & 0.6316        & 10290       & 3.116        & 0.1429     \\
                        & $s$         & 0.0754              & 0.0856        & 240.2       & 0.691        & 0.0374     \\ \hline
\multirow{3}{*}{C-BRIk} & $\tilde{x}$ & 0.7800              & 0.6352        & 10260       & 2.000        & 0.1521     \\
                        & $\bar{x}$   & 0.7912              & 0.6432        & 10260       & 2.139        & 0.1600     \\
                        & $s$         & 0.0664              & 0.0804        & 224.9       & 0.422        & 0.0396     \\ 

\end{tabular}
\end{center}
\caption{Summary statistics for Model 1. The median, mean and variance of 1000 independent datasets for the five performance evaluation measures are provided. Best medians and means are bold-faced.}
\label{model1estimates}
\end{table}

Notably, in this model the variability of the first four measures is remarkably smaller for FABRIk. Here, the synthetic groups 2 and 3 are easily confounded and inappropriate initial seeds lead $k$-Means to merge these two groups into a single cluster and, consequently, to split one of the other groups into two clusters. This situation considerably reduces the ARI of the corresponding algorithms, whose distributions become bimodal.
As an example, we considered a single dataset following Model 1, and compared the output of FABRIk, with $B=25$, versus 25 runs of $k$-Means with random initialization. Our method correctly allocates all the elements ($ARI= 1$), whereas none of 25 runs of standard $k$-Means is capable of retrieving the correct grouping (average $ARI= 0.9177$), with the confusion of clusters 2 and 3 in four of the runs.

\begin{figure}
  \begin{center}
    \includegraphics[width=\textwidth, trim=0 1mm 0 0, clip]{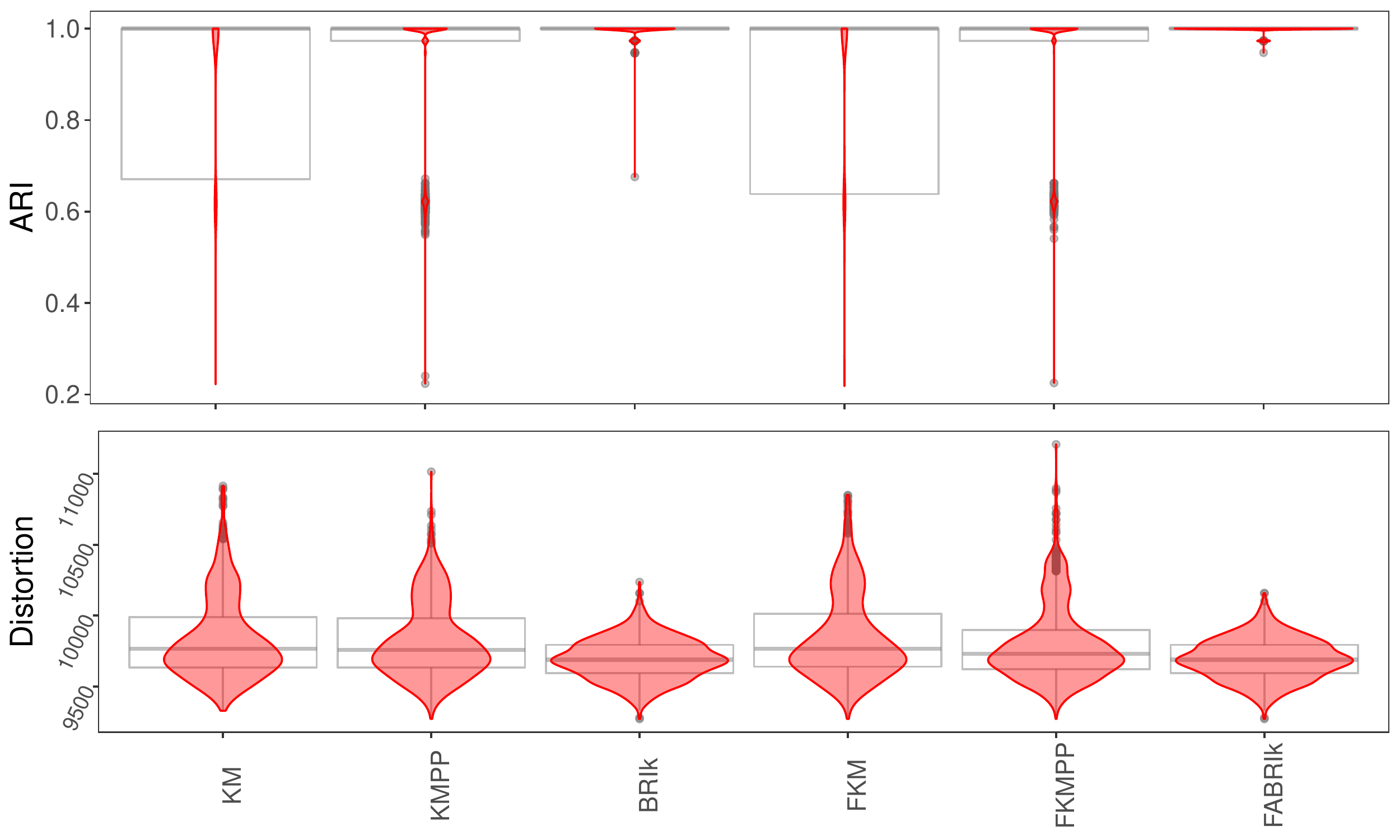}
    \caption{Violin plots, in red, for the distribution of the ARI index (top) and the distortion (bottom) in Model 1, with $\sigma=1$. The corresponding wider boxplots are superimposed in gray. BRIk and FABRIk have a consistent behaviour whereas the other methods have more spread or bimodal distributions, with heavy lower and upper tails, respectively.}
    \label{fig3}
  \end{center}
\end{figure}

Figure~\ref{fig3}, upper panel, depicts violin plots of the ARI distributions; the ones corresponding to the correctness (not shown) display a similar pattern.  The effect of wrong allocations is also reflected in the distribution of the distortion: all the methods except BRIk and FABRIk have bimodal densities or heavy upper tails, as shown in Figure~\ref{fig3}, bottom panel. This behavior is observed in a significant number of runs of the methods, but FABRIK --followed by BRIk-- seems to find the right partitioning more often. 

Despite a median test does not reject the equality of medians for the accuracy measures (correctness and ARI), we conducted pairwise t-tests for testing equality of means. As expected from their asymmetric distributions, we obtained p-values lower than $10^{-45}$ for the comparison of FABRIk with the other methods, except for BRIk, with p-values in the order of $10^{-3}$ and $10^{-6}$.

With respect to the missing data case (sparse data), we report in Table~\ref{model1missing} the results for a percentage $p=25\%$ of missing data. Similar relative outputs can be observed in the other cases. FABRIk is again the best method in terms of correctness, ARI and number of iterations to reach convergence, followed by BRIk and FKMPP. With respect to the distortion, FABRIk is only slightly surpassed by BRIk. Regarding the execution time, KM, BRIk and KMPP required longer times for the  interpolation step, and hence the methods based on FDA are, globally, a more suitable option.

\begin{table}[!htb]
\begin{center}
\begin{tabular}{ccccccc}
                        &             & Correctness     & ARI           & Distortion    & Iterations    & Exec. Time (s) \\ \hline
\multirow{3}{*}{KM}     & $\tilde{x}$ & 0.9900          & 0.9731        & 8607          & 3.000         & 0.2620         \\
                        & $\bar{x}$   & 0.9330          & 0.9152        & 8660          & 2.635         & 0.2693         \\
                        & $s$         & 0.1297          & 0.1459        & 257.9         & 0.608         & 0.0334         \\ \hline
\multirow{3}{*}{KMPP}   & $\tilde{x}$ & 0.9900          & 0.9731        & 8608          & 3.000         & 0.2653         \\
                        & $\bar{x}$   & 0.9383          & 0.9216        & 8650          & 2.564         & 0.2728         \\
                        & $s$         & 0.1254          & 0.1408        & 244.3         & 0.557         & 0.0335         \\ \hline
\multirow{3}{*}{BRIk}   & $\tilde{x}$ &\textbf{1.0000}  &\textbf{1.000} &\textbf{8572}  &\textbf{1.000} & 0.3569         \\
                        & $\bar{x}$   & 0.9938          & 0.9835        &\textbf{8571}  & 1.469         & 0.3662         \\
                        & $s$         & 0.2495          & 0.0210        & 152.1         & 0.509         & 0.0398         \\ \hline
\multirow{3}{*}{FKM}    & $\tilde{x}$ &\textbf{1.0000}  &\textbf{1.000} & 8615          & 2.000         &\textbf{0.1446} \\
                        & $\bar{x}$   & 0.9328          & 0.9193        & 8672          & 2.374         &\textbf{0.1526} \\
                        & $s$         & 0.1344          & 0.1489        & 258.9         & 0.528         & 0.0301         \\ \hline
\multirow{3}{*}{FKMPP}  & $\tilde{x}$ &\textbf{1.0000}  &\textbf{1.000} & 8611          & 2.000         & 0.1845         \\
                        & $\bar{x}$   & 0.9428          & 0.9307        & 8653          & 2.224         & 0.1882         \\
                        & $s$         & 0.1238          & 0.1358        & 240.5         & 0.498         & 0.0244         \\ \hline
\multirow{3}{*}{FABRIk} & $\tilde{x}$ &\textbf{1.0000}  &\textbf{1.000} & 8574          &\textbf{1.000} & 0.2757         \\
                        & $\bar{x}$   &\textbf{0.9957}  &\textbf{0.9886}& 8573          &\textbf{1.133} & 0.2807         \\
                        & $s$         & 0.2203          & 0.0185        & 152.3         & 0.340         & 0.0314         \\ 
\end{tabular}
\end{center}
\caption{Summary statistics for Model 1 with 25\% missing values. The median, mean and variance of 1000 independent datasets for the five performance evaluation measures are provided. Best medians and means are bold-faced.}
\label{model1missing}
\end{table}

%%%%%%%%%%%%%%%%%%%%%%%%%%%%%%%%%%
%%%%%%%%%    MODEL 2     %%%%%%%%%
%%%%%%%%%%%%%%%%%%%%%%%%%%%%%%%%%%

In contrast to the previous case, in Models 2--4 FABRIk has in general a distortion slightly higher than that of KM, BRIk and KMPP but smaller than that of the other initialization methods, as shown in Tables~\ref{model2statistics}--\ref{model4missing}.

This can be explained by accounting for the two different data spaces we are considering and our process of computing the distortion. From the point of view of the functional data space, FABRIk is simply the initialization of $k$-Means with appropriate seeds. However, from the point of view of the initial multivariate data space, the clustering obtained with FABRIk does not necessarily (and not even frequently) correspond to a local minimum of the distortion in this space, therefore yielding higher values of the objective function. Nevertheless, FABRIk is the best option among the functional methods. On the contrary, it consistently provides remarkably higher accuracy measures and faster convergence in terms of the number of iterations. It also has a longer execution time, as expected. However, its ranking improves again if missing data are considered.

To assess the relevance of this increment in the computational cost we compared these results with the strategy of initializing $k$-Means several times and choosing the set of seeds providing the lowest distortion. For instance, in Model 2, KM with 200 random starts increases the average ARI from 0.4200 to 0.4549, which is far from the 0.6467 obtained with our method, and requires 0.1263 seconds on average. 

In these models, all the pairwise tests for equality of means and medians for correctness, ARI and distortion, to compare FABRIk with the other methods, yielded p-values with an order of magnitude between $10^{-262}$ and $10^{-9}$. In summary, we can report a significant improvement over the alternatives.

\begin{table}[!htb]
\begin{center}
\begin{tabular}{ccccccc}
                        &             & Correctness           & ARI           & Distortion  & Iterations       & Exec. Time (s)     \\ \hline
\multirow{3}{*}{KM}     & $\tilde{x}$ & 0.6600                & 0.4294        &\textbf{9568}& 4.000            &\textbf{$\sim$0}     \\
                        & $\bar{x}$   & 0.6524                & 0.4200        &\textbf{9568}& 3.721            &\textbf{0.0010}     \\
                        & $s$         & 0.0709                & 0.0856        & 141.9       & 0.771            & 0.0026     \\ \hline
\multirow{3}{*}{KMPP}   & $\tilde{x}$ & 0.6600                & 0.4265        & 9684        & 4.000            & 0.0030     \\
                        & $\bar{x}$   & 0.6527                & 0.4193        & 9687        & 3.701            & 0.0040     \\
                        & $s$         & 0.0711                & 0.0864        & 144.1       & 0.737            & 0.0044     \\ \hline
\multirow{3}{*}{BRIk}   & $\tilde{x}$ & 0.6700                & 0.4369        & 9575        &\textbf{2.000}    & 0.0828     \\
                        & $\bar{x}$   & 0.6596                & 0.4315        & 9573        & 3.484            & 0.0885     \\
                        & $s$         & 0.0668                & 0.0853        & 142.4       & 0.818            & 0.0221     \\ \hline
\multirow{3}{*}{FKM}    & $\tilde{x}$ & 0.8100                & 0.6109        & 9659        & 3.000            & 0.1046     \\
                        & $\bar{x}$   & 0.7762                & 0.6113        & 9659        & 2.795            & 0.1068     \\
                        & $s$         & 0.0976                & 0.0882        & 142.6       & 0.594            & 0.0124     \\ \hline
\multirow{3}{*}{FKMPP}  & $\tilde{x}$ & 0.8000                & 0.6103        & 9656        & 3.000            & 0.1089     \\
                        & $\bar{x}$   & 0.7730                & 0.6099        & 9659        & 2.632            & 0.1140     \\
                        & $s$         & 0.0992                & 0.0903        & 142.8       & 0.599            & 0.0261     \\ \hline
\multirow{3}{*}{FABRIk} & $\tilde{x}$ &\textbf{0.8400}        &\textbf{0.6513}& 9650        &\textbf{2.000}    & 0.1862     \\
                        & $\bar{x}$   &\textbf{0.8253}        &\textbf{0.6467}& 9651        &\textbf{1.949}    & 0.1955     \\
                        & $s$         & 0.0606                & 0.0745        & 142.4       & 0.335            & 0.0385     \\ %\hline
\end{tabular}
\caption{Summary statistics for Model 2. The median, mean and variance of 1000 independent datasets for the five performance evaluation measures are provided. Best medians and means are bold-faced.}
\label{model2statistics}
\end{center}
\end{table}

%%%%%% missing data table
\begin{table}[!htb]
\begin{center}
\begin{tabular}{ccccccc}
                        &             & Correctness          & ARI           & Distortion  & Iterations   & Exec. Time (s) \\ \hline
\multirow{3}{*}{KM}     & $\tilde{x}$ & 0.6400               & 0.3825        & 8427        & 4.000        & 0.2170         \\
                        & $\bar{x}$   & 0.6380               & 0.3810        &\textbf{8429}& 3.653        & 0.2248         \\
                        & $s$         & 0.0674               & 0.0787        & 149.9       & 0.716        & 0.0312         \\ \hline
\multirow{3}{*}{KMPP}   & $\tilde{x}$ & 0.6400               & 0.3817        &\textbf{8426}& 4.000        & 0.2200         \\
                        & $\bar{x}$   & 0.6333               & 0.3782        &\textbf{8429}& 3.651        & 0.2288         \\
                        & $s$         & 0.0655               & 0.0749        & 149.1       & 0.740        & 0.0327        \\ \hline
\multirow{3}{*}{BRIk}   & $\tilde{x}$ & 0.6500               & 0.3973        & 8430        & 3.000        & 0.2974         \\
                        & $\bar{x}$   & 0.6434               & 0.3948        & 8432        & 3.336        & 0.3088         \\
                        & $s$         & 0.0594               & 0.0736        & 149.9       & 0.829        & 0.0381      \\ \hline
\multirow{3}{*}{FKM}    & $\tilde{x}$ & 0.7500               & 0.5245        & 8517        & 3.000        &\textbf{0.1101} \\
                        & $\bar{x}$   & 0.7344               & 0.5270        & 8514        & 2.937        &\textbf{0.1110} \\
                        & $s$         & 0.0829               & 0.0776        & 153.1       & 0.661        & 0.0122         \\ \hline
\multirow{3}{*}{FKMPP}  & $\tilde{x}$ & 0.7400               & 0.5182        & 8518        & 3.000        &0.1108 \\
                        & $\bar{x}$   & 0.7260               & 0.5219        & 8516        & 2.776        &0.1164 \\
                        & $s$         & 0.0864               & 0.0785        & 152.5       & 0.615        & 0.0218         \\ \hline
\multirow{3}{*}{FABRIk} & $\tilde{x}$ &\textbf{0.7800}       &\textbf{0.5537}& 8513        &\textbf{2.000}& 0.1896         \\
                        & $\bar{x}$   &\textbf{0.7690}       &\textbf{0.5516}& 8508        & 2.001        & 0.1967         \\
                        & $s$         & 0.0669               & 0.0753        & 151.3       & 0.324        & 0.0306        \\ %\hline
\end{tabular}
\end{center}
\caption{Summary statistics for Model 2 with 25\% missing values. The median, mean and variance of 1000 independent datasets for the five performance evaluation measures are provided. Best medians and means are bold-faced.}
\label{model2missing}
\end{table}

%%%%%%%%%%%%%%%%%%%%%%%%%%%%%%%%%%
%%%%%%%%%    MODEL 3     %%%%%%%%%
%%%%%%%%%%%%%%%%%%%%%%%%%%%%%%%%%%

\begin{table}[!htb]
\begin{center}
\begin{tabular}{ccccccc}
                        &             & Correctness     & ARI           & Distortion   & Iterations   & Exec. Time (s)  \\ \hline
\multirow{3}{*}{KM}     & $\tilde{x}$ & 0.4400          & 0.2151        &\textbf{23680}& 4.000        &\textbf{0.0026} \\
                        & $\bar{x}$   & 0.4408          & 0.2176        &\textbf{23680}& 4.116        &\textbf{0.0025}  \\
                        & $s$         & 0.0363          & 0.0377        & 228.4        & 0.853        & 0.0021          \\ \hline
\multirow{3}{*}{KMPP}   & $\tilde{x}$ & 0.4400          & 0.2123        &\textbf{23680}& 4.000        & 0.0100          \\
                        & $\bar{x}$   & 0.4396          & 0.2150        &\textbf{23680}& 4.113        & 0.0132          \\
                        & $s$         & 0.0355          & 0.0360        & 227.1        & 0.860        & 0.0042          \\ \hline
\multirow{3}{*}{BRIk}   & $\tilde{x}$ & 0.4320          & 0.2424        &\textbf{23680}& 4.000        & 0.2445          \\
                        & $\bar{x}$   & 0.4361          & 0.2430        & 23690        & 4.169        & 0.2484          \\
                        & $s$         & 0.0324          & 0.0396        & 225.3        & 0.876        & 0.0221          \\ \hline
\multirow{3}{*}{FKM}    & $\tilde{x}$ & 0.5200          & 0.3190        & 23980        &\textbf{3.000}& 0.2065          \\
                        & $\bar{x}$   & 0.5203          & 0.3183        & 23980        & 3.472        & 0.2110          \\
                        & $s$         & 0.0566          & 0.0484        & 230.6        & 0.693        & 0.0197          \\ \hline
\multirow{3}{*}{FKMPP}  & $\tilde{x}$ & 0.5200          & 0.3106        & 23980        &\textbf{3.000}& 0.2158          \\
                        & $\bar{x}$   & 0.5191          & 0.3118        & 23970        & 3.433        & 0.2203          \\
                        & $s$         & 0.0562          & 0.0481        & 230.4        & 0.684        & 0.0203          \\ \hline
\multirow{3}{*}{FABRIk} & $\tilde{x}$ &\textbf{0.5280}  &\textbf{0.3285}& 23970        &\textbf{3.000}& 0.4350          \\
                        & $\bar{x}$   &\textbf{0.5314}  &\textbf{0.3283}& 23970        &\textbf{2.603}& 0.4415          \\
                        & $s$         & 0.0559          & 0.0454        & 229.6        & 0.556        & 0.0278          \\ 
\end{tabular}
\end{center}
\caption{Summary statistics for Model 3. The median, mean and variance of 1000 independent datasets for the five performance evaluation measures are provided. Best medians and means are bold-faced.}
\label{model3estimates}
\end{table}

\begin{table}[!htb]
\begin{center}
\begin{tabular}{ccccccc}
                        &             & Correctness    & ARI             & Distortion     & Iterations     & Exec. Time (s)  \\ \hline
\multirow{3}{*}{KM}     & $\tilde{x}$ & 0.4320         & 0.2010          & \textbf{20850} & 4.000          & 0.6744          \\
                        & $\bar{x}$   & 0.4363         & 0.2030          & \textbf{20850} & 4.091          & 0.6870          \\
                        & $s$         & 0.0368         & 0.0363          & 236.4          & 0.807          & 0.0530          \\ \hline
\multirow{3}{*}{KMPP}   & $\tilde{x}$ & 0.4320         & 0.1969          & \textbf{20850} & 4.000          & 0.6835          \\
                        & $\bar{x}$   & 0.4356         & 0.2004          & \textbf{20850} & 4.093          & 0.6960          \\
                        & $s$         & 0.0358         & 0.0367          & 234.6          & 0.813          & 0.0532          \\ \hline
\multirow{3}{*}{BRIk}   & $\tilde{x}$ & 0.4320         & 0.2221          & 20860          & 4.000          & 0.9314          \\
                        & $\bar{x}$   & 0.4340         & 0.2235          & 20860          & 4.108          & 0.9473          \\
                        & $s$         & 0.0347         & 0.0399          & 237.3          & 0.883          & 0.0648          \\ \hline
\multirow{3}{*}{FKM}    & $\tilde{x}$ & \textbf{0.4880}& 0.2623          & 21100          & \textbf{3.000} & \textbf{0.2121} \\
                        & $\bar{x}$   & 0.4894         & 0.2654          & 21100          & 3.574          & \textbf{0.2187} \\
                        & $s$         & 0.0490         & 0.0443          & 240.7          & 0.720          & 0.0238          \\ \hline
\multirow{3}{*}{FKMPP}  & $\tilde{x}$ & \textbf{0.4880}& 0.2629          & 21100          & \textbf{3.000} & 0.2202          \\
                        & $\bar{x}$   & 0.4912         & 0.2680          & 21100          & 3.543          & 0.2268          \\
                        & $s$         & 0.0506         & 0.0443          & 239.6          & 0.706          & 0.0227          \\ \hline
\multirow{3}{*}{FABRIk} & $\tilde{x}$ & \textbf{0.4880}& \textbf{0.2785} & 21090          & \textbf{3.000} & 0.4511          \\
                        & $\bar{x}$   & \textbf{0.4946}& \textbf{0.2794} & 21090          & \textbf{2.698} & 0.4620          \\
                        & $s$         & 0.0463         & 0.0421          & 239.4          & 0.694          & 0.0370          \\ 
\end{tabular}
\end{center}
\caption{Summary statistics for Model 3 with 25\% missing values. The median, mean and variance of 1000 independent datasets for the five performance evaluation measures are provided. Best medians and means are bold-faced.}
\label{model3missing}
\end{table}

%%%%%%%%%%%%%%%%%%%%%%%%%%%%%%%%%%
%%%%%%%%%    MODEL 4     %%%%%%%%%
%%%%%%%%%%%%%%%%%%%%%%%%%%%%%%%%%%

\begin{table}[!htb]
\begin{center}
\begin{tabular}{ccccccc}
                        &             & Correctness        & ARI           & Distortion  & Iterations   & Exec. Time (s)   \\ \hline
\multirow{3}{*}{KM}     & $\tilde{x}$ & 0.5700             & 0.3070        & 1894        & 3.000        &\textbf{$\sim$0}  \\
                        & $\bar{x}$   & 0.5692             & 0.3071        & 1894        & 3.394        &\textbf{0.0003}   \\
                        & $s$         & 0.0537             & 0.0558        & 61.26       & 0.682        & 0.0012            \\ \hline
\multirow{3}{*}{KMPP}   & $\tilde{x}$ & 0.5800             & 0.3123        & 1895        & 3.000        & 0.0016            \\
                        & $\bar{x}$   & 0.5719             & 0.3089        & 1894        & 3.364        & 0.0021            \\
                        & $s$         & 0.0572             & 0.0602        & 60.78       & 0.716        & 0.0028            \\ \hline
\multirow{3}{*}{BRIk}   & $\tilde{x}$ & 0.5800             & 0.3176        &\textbf{1891}& 3.000        & 0.0208            \\
                        & $\bar{x}$   & 0.5806             & 0.3183        &\textbf{1891}& 2.667        & 0.0227            \\
                        & $s$         & 0.0525             & 0.0568        & 61.18       & 0.682        & 0.0091            \\ \hline
\multirow{3}{*}{FKM}    & $\tilde{x}$ & 0.6200             & 0.3678        & 1938        & 3.000        & 0.1207            \\
                        & $\bar{x}$   & 0.6147             & 0.3658        & 1937        & 2.992        & 0.1239            \\
                        & $s$         & 0.0538             & 0.0594        & 63.27       & 0.696        & 0.0180            \\ \hline
\multirow{3}{*}{FKMPP}  & $\tilde{x}$ & 0.6200             & 0.3657        & 1940        & 3.000        & 0.0932            \\
                        & $\bar{x}$   & 0.6110             & 0.3643        & 1938        & 2.902        & 0.0977            \\
                        & $s$         & 0.0553             & 0.0616        & 62.72       & 0.692        & 0.0188            \\ \hline
\multirow{3}{*}{FABRIk} & $\tilde{x}$ &\textbf{0.6300}     &\textbf{0.3730}& 1935        &\textbf{2.000}& 0.1128            \\
                        & $\bar{x}$   &\textbf{0.6293}     &\textbf{0.3772}& 1933        &\textbf{2.092}& 0.1178            \\
                        & $s$         & 0.0448             & 0.0560        & 63.30       & 0.368        & 0.0203            \\ 
\end{tabular}
\end{center}
\caption{Summary statistics for Model 4. The median, mean and variance of 1000 independent datasets for the five performance evaluation measures are provided. Best medians and means are bold-faced.}
\label{model4estimates}
\end{table}

\begin{table}[!htb]
\begin{center}
\begin{tabular}{ccccccc}
                        &             & Correctness             & ARI           & Distortion  & Iterations   & Exec. Time (s)   \\ \hline
\multirow{3}{*}{KM}     & $\tilde{x}$ & 0.5700                  & 0.2858        & 1653        & 3.000        &\textbf{0.0432}   \\
                        & $\bar{x}$   & 0.5634                  & 0.2843        & 1654        & 3.363        &\textbf{0.0405}   \\
                        & $s$         & 0.0528                  & 0.0567        & 62.52       & 0.688        & 0.01197          \\ \hline
\multirow{3}{*}{KMPP}   & $\tilde{x}$ & 0.5600                  & 0.2813        & 1654        & 3.000        & 0.0461           \\
                        & $\bar{x}$   & 0.5626                  & 0.2818        & 1655        & 3.329        & 0.0421           \\
                        & $s$         & 0.0518                  & 0.0561        & 62.00       & 0.661        & 0.0129           \\ \hline
\multirow{3}{*}{BRIk}   & $\tilde{x}$ & 0.5700                  & 0.2896        &\textbf{1650}&\textbf{2.000}& 0.06248          \\
                        & $\bar{x}$   & 0.5691                  & 0.2897        &\textbf{1651}& 2.540        & 0.0616           \\
                        & $s$         & 0.0460                  & 0.0539        & 61.90       & 0.655        & 0.0141           \\ \hline
\multirow{3}{*}{FKM}    & $\tilde{x}$ & 0.5900                  & 0.3119        & 1686        & 3.000        & 0.1211           \\
                        & $\bar{x}$   & 0.5837                  & 0.3122        & 1687        & 3.043        & 0.1248           \\
                        & $s$         & 0.0503                  & 0.0584        & 64.06       & 0.717        & 0.0185           \\ \hline
\multirow{3}{*}{FKMPP}  & $\tilde{x}$ & 0.5800                  & 0.3082        & 1687        & 3.000        & 0.0937           \\
                        & $\bar{x}$   & 0.5802                  & 0.3085        & 1688        & 2.981        & 0.0938           \\
                        & $s$         & 0.0552                  & 0.0618        & 63.96       & 0.675        & 0.0160           \\ \hline
\multirow{3}{*}{FABRIk} & $\tilde{x}$ &\textbf{0.6000}          &\textbf{0.3238}& 1682        &\textbf{2.000}& 0.1094           \\
                        & $\bar{x}$   &\textbf{0.5959}          &\textbf{0.3205}& 1683        &\textbf{2.119}& 0.1135           \\
                        & $s$         & 0.0456                  & 0.0567        & 63.48       & 0.386        & 0.0190           \\ 
\end{tabular}
\end{center}
\caption{Summary statistics for Model 4 with 25\% missing values. The median, mean and variance of 1000 independent datasets for the five performance evaluation measures are provided. Best medians and means are bold-faced.}
\label{model4missing}
\end{table}

%%%%%%%%%%%%%%%%%%%%%%%%%%%%%%%%%%%%%%%%%%%
%%%%%%          2. REAL DATA        %%%%%%% 
%%%%%%%%%%%%%%%%%%%%%%%%%%%%%%%%%%%%%%%%%%%

\paragraph{Real data} We next applied all the initialization methods to the real  data.

%%%%   ECG
For the ECG dataset, the DFs were set to 15 according to the elbow rule and we chose an oversampling factor of 1 for speed, as using a denser time grid produces a similar output. Table~\ref{ECG200} summarizes our results. Note that the quality of the clustering recovery in terms of ARI is small. We do not find prominent differences across methods. In particular, all of them require a single iteration to converge and have the same median for correctness, ARI and distortion. Yet, FABRIk leads to the best average correctness and second best average ARI and distortion, and has the smallest standard deviations. This corresponds to a single-mode distribution: a scenario similar to that depicted in Figure~\ref{fig3} for simulated data.  As usual, FABRIk is largely outperformed in terms of execution time by those methods that do not rely on the B-spline approximation.

\begin{table}[!htb]
\begin{center}

\begin{tabular}{ccccccc}
                        &             & Correctness        & ARI           & Distortion        & Iterations   & Exec. Time (s)    \\ \hline
\multirow{3}{*}{KM}     & $\tilde{x}$ & \textbf{0.7450}    &\textbf{0.2194}& \textbf{5117}     &\textbf{1.000}&\textbf{0.0010}    \\
                        & $\bar{x}$   & 0.7307             & 0.1877        & 5370              &\textbf{1.000}&\textbf{0.0008}    \\
                        & $s$         & 0.0209             & 0.0465        & 371.6             & 0.000        & 0.0009 \\ \hline
\multirow{3}{*}{KMPP}   & $\tilde{x}$ & \textbf{0.7450}    &\textbf{0.2194}& \textbf{5117}     &\textbf{1.000}& 0.0020          \\
                        & $\bar{x}$   & 0.7278             & 0.1812        & 5422              &\textbf{1.000}& 0.0024         \\
                        & $s$         & 0.0219             & 0.0485        & 388.0             & 0.000        & 0.0022          \\ \hline
\multirow{3}{*}{BRIk}   & $\tilde{x}$ & \textbf{0.7450}    &\textbf{0.2194}& \textbf{5117}     &\textbf{1.000}& 0.0509           \\
                        & $\bar{x}$   & \textbf{0.7374}    &\textbf{0.2025}& \textbf{5252}     &\textbf{1.000}& 0.0554           \\
                        & $s$         & 0.0168             & 0.0374        & 299.3             & 0.000        & 0.0167        \\ \hline
\multirow{3}{*}{FKM}    & $\tilde{x}$ & \textbf{0.7450}             &\textbf{0.2194}&\textbf{5117}      &\textbf{1.000}& 0.3850      \\
                        & $\bar{x}$   & 0.7326             & 0.1908        & 5364              &\textbf{1.000}& 0.3860      \\
                        & $s$         & 0.0184             & 0.0428        & 370.0             & 0.000        & 0.0014       \\ \hline
\multirow{3}{*}{FKMPP}  & $\tilde{x}$ & \textbf{0.7450}    &\textbf{0.2194}& \textbf{5117}     &\textbf{1.000}& 0.3871       \\
                        & $\bar{x}$   & 0.7304             & 0.1855        & 5410              &\textbf{1.000}& 0.3875     \\
                        & $s$         & 0.0193             & 0.0446        & 385.7             & 0.000        & 0.0040    \\ \hline
\multirow{3}{*}{FABRIk} & $\tilde{x}$ & \textbf{0.7450}    &\textbf{0.2194}& \textbf{5117}     &\textbf{1.000}& 0.4459         \\
                        & $\bar{x}$   & \textbf{0.7374}    & 0.2018        & 5269              &\textbf{1.000}& 0.4484        \\
                        & $s$         & 0.0157             & 0.0363        & 314.1             & 0.000        & 0.0129       \\ 
\end{tabular}

\end{center}
\caption{Summary statistics for the ECG dataset. The median, mean and variance of 1000 runs of each initialization method for the five performance evaluation measures are provided. Best medians and means are bold-faced.}
\label{ECG200}
\end{table}

%%%%   Gyroscope
For the Gyroscope dataset, the DFs were set to 15, once more, according to the elbow rule. The oversampling factor was set to 1. Again, a similar performance is observed for higher values of this parameter, which does not influence the final results.

\begin{table}[!htb]
\begin{center}
\begin{tabular}{ccccccc}
                        &             & Correctness         & ARI              & Distortion     & Iterations         & Exec. Time (s)    \\ \hline
\multirow{3}{*}{KM}     & $\tilde{x}$ & 0.6591              & 0.6154           & 4645           & \textbf{2.000}     &\textbf{0.0010} \\
                        & $\bar{x}$   & 0.7565              & 0.7126           & 4717           & 2.037              &\textbf{0.0014} \\
                        & $s$         & 0.1726              & 0.1830           & 636.8          & 0.334              & 0.0007         \\ \hline
\multirow{3}{*}{KMPP}   & $\tilde{x}$ & 0.6591              & 0.6154           & \textbf{3929}  & \textbf{2.000}     & 0.0050          \\
                        & $\bar{x}$   & 0.6887              & 0.6387           & 4483           & 1.899              & 0.0055          \\
                        & $s$         & 0.1535              & 0.1729           & 698.2          & 0.373              & 0.0019          \\ \hline
\multirow{3}{*}{BRIk}   & $\tilde{x}$ & 0.6591              & 0.6154           & \textbf{3929}  & \textbf{2.000}     & 0.4644           \\
                        & $\bar{x}$   & 0.8067              & 0.7650           & \textbf{4261}  & \textbf{1.570}     & 0.4724           \\
                        & $s$         & 0.1588              & 0.1609           & 356.9          & 0.495              & 0.0582           \\ \hline
\multirow{3}{*}{FKM}    & $\tilde{x}$ & 0.6591              & 0.6154           & 4645           & \textbf{2.000}     & 0.2509         \\
                        & $\bar{x}$   & 0.7727              & 0.7291           & 4753           & 2.003              & 0.2514         \\
                        & $s$         & 0.1810              & 0.1920           & 610.1          & 0.345              & 0.0008         \\ \hline
\multirow{3}{*}{FKMPP}  & $\tilde{x}$ & 0.6591              & 0.6154           & \textbf{3929}  & \textbf{2.000}     & 0.2549          \\
                        & $\bar{x}$   & 0.7090              & 0.6615           & 4429           & 1.820              & 0.2554          \\
                        & $s$         & 0.1541              & 0.1688           & 636.9          & 0.426              & 0.0016          \\ \hline
\multirow{3}{*}{FABRIk} & $\tilde{x}$ & \textbf{0.9773}     & \textbf{0.9379}  & 4645           & \textbf{2.000}     & 0.7123           \\
                        & $\bar{x}$   & \textbf{0.8471}     & \textbf{0.8060}  & 4352           & 1.613              & 0.7190           \\
                        & $s$         & 0.1565              & 0.1586           & 351.9          & 0.487              & 0.0556           \\ 
\end{tabular}
\end{center}
\caption{Summary statistics for the Gyroscope dataset. The median, mean and variance of 1000 runs of each initialization method for the five performance evaluation measures are provided. Best medians and means are bold-faced.}
\label{Gyroscope}
\end{table}

In contrast to the previous case, the values of correctness and ARI are much higher. However the FABRIk method finds more accurate groups, obtaining ARI values larger than 0.9 in roughly 60\% of the iterations, whereas for instance, this percentage is around 35\% and 20\% for KM and FKMPP, respectively. Also, for distortion it is the second best option (after BRIk) and shows the least variability, followed by BRIk. In fact, for these two methods, the accuracy and dispersion measures have bi-modal distributions, while those corresponding to the other algorithms present three or more peaks. With respect to the number of iterations all methods have similar values, with BRIk and FABRIk slightly better on average. However, the computational cost of our method, is the largest one.

\subsection{Implementation}
\begin{figure}
  \begin{center}
    \includegraphics[width=\textwidth]{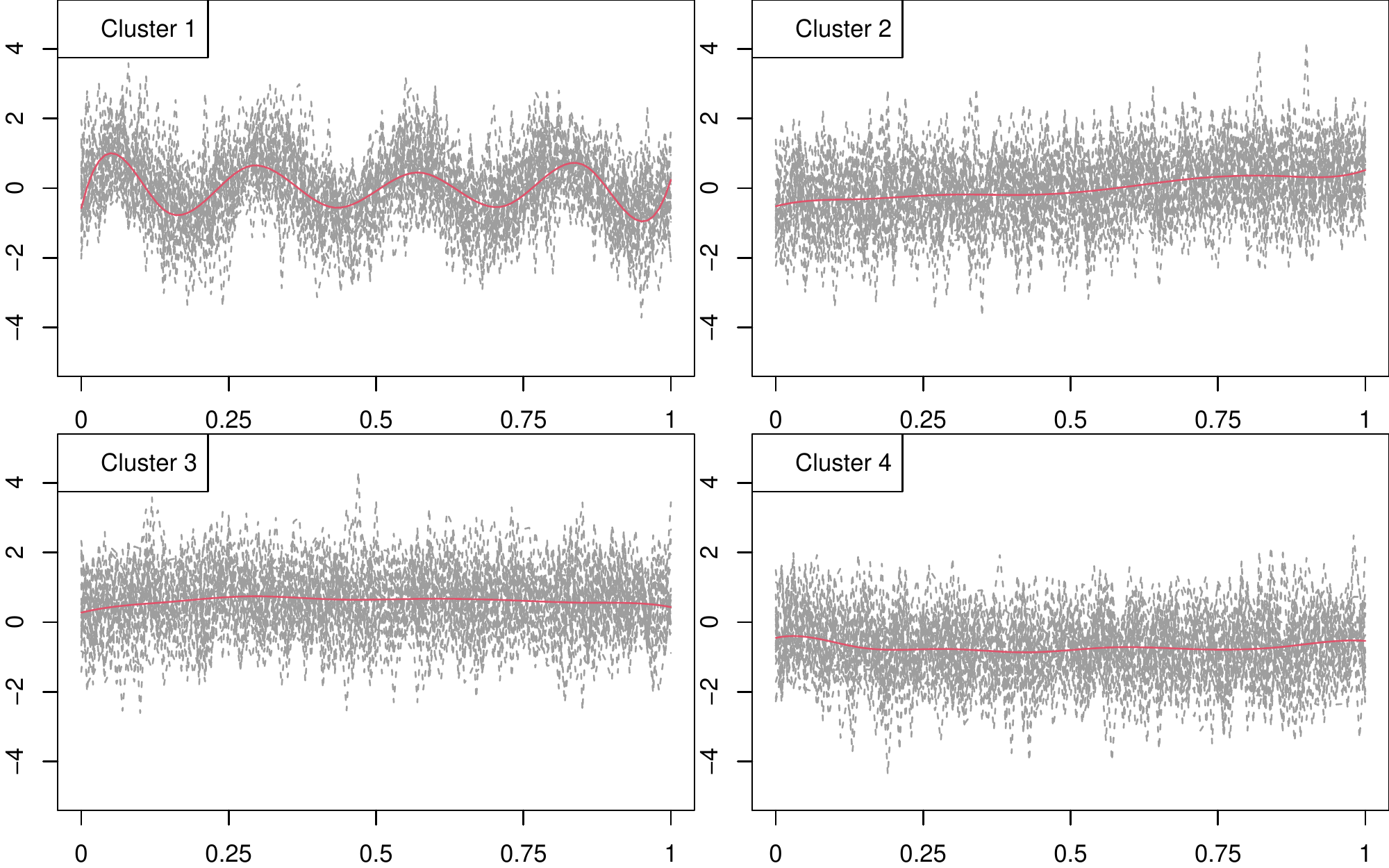}
    \caption{Representation of the four clusters retrieved by FABRIk for a dataset following Model 1 with $\sigma = 1$, along with the final centroid (solid line). Our method correctly allocates all the elements ($ARI=1$).     }
    \label{fig4}
  \end{center}
\end{figure}

We have implemented an R package, \verb|briKmeans|, to provide the basic tools to run both BRIk and FABRIk. Users can tune the different parameters of the methods through the functions parameters and retrieve the corresponding initial seeds and the resulting $k$-Means output, which includes the partitioning of the data set. For instance, the following simple call\\ \verb|> fabrik(exampleM1, k=4, degFr=10)|\\
will run FABRIk with DFs set to 10 and the rest of parameters set to default, and return \verb|k=4| clusters for the dataset \verb|exampleM1|. The clusters can be visualized individually in parallel coordinates \citep{Inselberg:85} by means of the \verb|plotKmeansClustering| function, including the final centroids. In Figure~\ref{fig4} we illustrate this representation for a dataset following Model 1, with $\sigma=1$. Note that users can also turn to the \verb|elbowRule| function to plot the distortion associated to FABRIk against the DFs in order to optimize this parameter.

%%%%%%%%%%%%%%%%%%%%%%%%%%%%%%%%%
%%%%%        CONCLUSION   
%%%%%%%%%%%%%%%%%%%%%%%%%%%%%%%%%%

\section{Conclusion}\label{section.discussion}

\hyphenation{si-mu-la-ted} \hyphenation{re-gar-ding}

In this work we have developed FABRIk, an initialization method for \textit{k}-Means that extends the BRIk algorithm to the functional data case. It takes $d$-dimensional longitudinal observations from continuous functions as an input dataset and returns the $D$-dimensional initial seeds for $k$-Means after a functional approximation process via B-splines and a re-sampling stage. 

Similarly to its precursor BRIk, our method is flexible in several ways. The number of bootstrap replicates $B$ can be tuned by the user; in general, low values of $B$ are enough to produce a relevant improvement over the alternatives. Additionally, the DFs and the oversampling factor $m$ can be chosen to best adapt to the data. An oversampling factor of 1 has proven to yield similar results to higher values of this parameter, while remaining less computationally expensive. In particular, the DFs are selected according to the elbow rule. Nevertheless, our experiments show that a wide range of values for these parameters are also suitable. The clustering algorithm used to partition the cluster centers is an extra feature that can be determined by the user. Finally, one could potentially use any feasible data depth definition, but our recommendation is to choose MBD for its fast computation, its applicability to both functional and multivariate data and because it has proven to score high in the accuracy measures.

We have made both methods publicly available through the R package \verb|briKmeans| (on the CRAN repository), which also allows following the elbow-like rule for selecting suitable values of the DF parameter and representing the clusters, along with the final centroids, in parallel coordinates.

 We have compared our functional initialization strategy to its multivariate version and to two more techniques, with and  without the FDA approach. Furthermore, we have assessed the behavior of the methods based on clustering the B-splines coefficients obtained for each data point, which have proven to be poor competitors.

Generally speaking, FABRIk works well with both synthetic and real data. It is an advantageous method that offers higher quality in terms of clustering recovery at the cost of a longer computational time and, commonly, a slightly larger distortion. In addition, we have shown that in some situations,  particularly with the real data we have considered, FABRIk rises as a more reliable way of initializing $k$-Means, which consistently provides better accuracy results with lower variance. Moreover, as any technique based on a functional approximation of the observations, it allows denoising and imputation of missing data.

%%%%%%%%%%%%%%%%%%%%%%%%%%%%%%%%%
%%%%%     ACKNOWLEDGEMENTS 
%%%%%%%%%%%%%%%%%%%%%%%%%%%%%%%%%

\section*{Acknowledgements}

\paragraph{Funding} This work was partially supported by the Spanish Ministry of Education [collaboration grant in university departments, Archive ID 18C01/003730] and the Spanish Ministry of Science, Innovation and Universities [grants numbers FIS2017-84440-C2-2-P and MTM2017-84446-C2-2-R].

%%%%%%%%%%%%%%%%%%%%%%%%%%%%%%%%%
%%%%%     BIBLIOGRAPHY
%%%%%%%%%%%%%%%%%%%%%%%%%%%%%%%%%

%\printbibliography

%\bibliography{myBib.bib}

\end{document}